\documentclass[12pt]{article}
\usepackage{hyperref}
\usepackage{amsmath,amssymb}
\usepackage{graphicx}
\usepackage{color}
\definecolor{myblue}{rgb}{0.14,0.11,0.49}
\definecolor{myred}{rgb}{0.74,0.22,0.15}
\definecolor{mygreen}{rgb}{0.05,0.52,0.42}
\definecolor{myyellow}{rgb}{0.96,0.92,0.13}
\definecolor{myorange}{rgb}{1,0.61,0.36}
\definecolor{mypurple}{rgb}{0.71,0.02,1}
\definecolor{noir}{gray}{0.} 

\definecolor{htc}{rgb}{1,1,1} 

\newcommand{\Mat}[1]{{{\boldsymbol{#1}}}}
\newcommand{\abs}[1]{\left\vert#1\right\vert}
\def\be{\begin{equation}}
\def\ee{\end{equation}}
\def\bea{\begin{eqnarray}}
\def\eea{\end{eqnarray}}
\def\bc{\begin{center}}
\def\ec{\end{center}}
\def\bi{\begin{itemize}}
\def\ei{\end{itemize}}
\def\bs{\begin{slide}}
\def\es{\end{slide}}
\def\dd{\mathrm{d}}

\def\noi{\noindent}
\title{{\bf Charge conservation in a gravitational field in the scalar ether theory} \footnote{
This work has been published in Open Physics, Vol. 15, pp. 877-890 (2017), DOI: \href{https://doi.org/10.1515/phys-2017-0105}{10.1515/phys-2017-0105} (Open Access). It is licensed under the Creative Commons Attribution-NonCommercial-NoDerivatives 4.0 License.
}}
\author{
Mayeul Arminjon\\
\small\it Univ. Grenoble Alpes, CNRS, Grenoble INP 
, 3SR, F-38000 Grenoble, France
} 
\date{}
			     
\begin{document}

\maketitle

\begin{abstract}

\noi A modification of the Maxwell equations due to the presence of a gravitational field was formerly proposed for a scalar theory with a preferred reference frame. With this modification, the electric charge is not conserved. The aim of the present work was to numerically assess the amount of charge production or destruction. We propose an asymptotic scheme for the electromagnetic field in a weak and slowly varying gravitational field. This scheme is valid independently of the theory and the ``gravitationally-modified" Maxwell equations. Then we apply this scheme to plane waves and to a group of Hertzian dipoles in the scalar ether theory. The predicted amounts of charge production/destruction discard the formerly proposed gravitationally-modified Maxwell equations. The theoretical reason for that is the assumption that the total energy tensor is the sum of the energy tensor of the medium producing the electromagnetic (e.m.) field and the e.m. energy tensor. This means that an additional, ``interaction" tensor has to be present. With this assumption, the standard Maxwell equations in a curved spacetime, which predict charge conservation, are compatible with the investigated theory. We find that the interaction energy might contribute to the dark matter.\\

\end{abstract}


\section{Introduction and summary}\label{Intro}

Since the standard Maxwell equations apply in special relativity i.e. in a flat Minkowski spacetime, any theory of gravitation 
with a curved spacetime has to modify them, while ensuring that the modification reduces to the standard form in the particular 
case of flat spacetime. In a foregoing work \cite{A54}, such a modification of the Maxwell equations in a gravitational field 
has been proposed for a scalar theory with a preferred reference frame, or (in short) with an ``ether". (That theory is 
compatible with special relativity, however, and has passed a number of tests; see the Introduction of Ref. \cite{A54}, where the motivations for the theory are also displayed.) It turns out that this modification leads to a violation of charge conservation if the gravitational field is time-dependent in the preferred frame \cite{A54}. However, the {\it amount} of that violation, which is the really important thing, could not be assessed at that stage. \\

The purpose of the present work, therefore, was to provide such assessment on a solid basis. To begin with, Section \ref{Main} summarizes the equations for the electromagnetic (e.m.) field in the theory, as given in Ref. \cite{A54}. Then Section \ref{Non-conserv-exact} establishes the exact equations for charge production/destruction which result from those equations. In Section \ref{PN}, we build an asymptotic approximation scheme for the electromagnetic field in a weak and slowly-varying gravitational field. This scheme has a general value, its main equations being valid for quite any theory of gravitation with a curved spacetime and the corresponding modification of the Maxwell equations. By using this scheme, in Sect. \ref{Explicit} we obtain an explicit expression for the charge production rate that results from the investigated modification of the Maxwell equations in a gravitational field. We assess the time variation of the Newtonian potential and gravity acceleration, which enter that expression. We apply this to a plane wave in Sect. \ref{Plane}, and to a group of Hertzian dipoles in Sect. \ref{Hertz}, thus getting figures. To do that, we follow a domain close to the source of the e.m. field in its motion through the ``ether" of the theory and we integrate the charge production rate in small subdomains, using Lorentz transformations forth and back between the moving frame and the ether. It turns out that, for a reasonable velocity through the ether, the amounts thus assessed seem much too high to be compatible with the experimental facts. \\

In Sects. \ref{Reason} and \ref{Solution}, we find the theoretical reason for this failure of the formerly proposed modification 
of the Maxwell equations in the investigated theory of gravitation: it is the assumption that the sum of two energy tensors, 
that of the charged medium and that of the e.m. field, is the total energy tensor which obeys the dynamical equation in a 
gravitational field. This assumption indeed leads to theoretical conclusions which cannot be generally true. Thus the only 
solution to this problem is to abandon this assumption, which means to introduce an additional, ``interaction" tensor (Sect. 
\ref{Solution}). Then, the Maxwell equations in a gravitational field are not determined any more by the equation for continuum 
dynamics satisfied by the charged medium subjected to the Lorentz force. In particular, one may assume the standard 
gravitationally-modified Maxwell equations used in general relativity, which, we show in Appendix \ref{Maxwell GR mean}, have a 
simple interpretation in the framework of the investigated theory. Section \ref{Solution} ends by showing that, very generally, 
the gravitationally-modified Maxwell equations remain compatible with the geometrical optics of the theory, i.e., with the 
dynamics of a photon subjected to a gravitational and a non-gravitational force, which is detailed in Appendix \ref{DynPhot}. 
Section \ref{Conclu} discusses the meaning of our findings. In particular, we suggest that the ``interaction" tensor, which we 
find to be necessarily present according to the present theory, should at least contribute to the ``missing mass" which has been 
invoked to explain motion at a galactic scale.

\section{Main equations for the e.m. field in the scalar theory}\label{Main}

In Ref. \cite{A54}, the equations have been written in the Gauss system of units, for simplicity. However, due to the fact that in the Gauss units the Coulomb law is written as
\be\label{Coulomb}
{\bf F} = q\,q'\frac{{\bf r}}{r^3},
\ee
in these units the charge dimension is $Q = (M\,L^3\,T ^{-2})^\frac{1}{2}$, where $M$, $L$ and $T $ represent mass, length, and time, respectively. This would introduce an undesirable coupling when writing weak-field asymptotic expansions, for in fact charge varies independently of mass, length, and time. Therefore, while presenting now the main equations \cite{A54}, we rewrite them in the so-called SI units, i.e., in the ``meter, kilogram, second, Amp\`ere" (MKSA) system. The electromagnetic field is defined by the antisymmetric space-time tensor $\Mat{F}$: $F_{\mu \nu 
} = - F_{\nu \mu \, }$, that obeys the standard first group of the Maxwell equations:
\be\label{Maxwell 1}
F_{\lambda \mu \, ,\nu } + F_{\mu \nu ,\lambda 
} + F_{\nu \lambda ,\mu } = F_{\lambda \mu \, ;\nu } + F_{\mu \nu ;\lambda 
} + F_{\nu \lambda ;\mu } = 0.
\ee 
The expression of the Lorentz force is 
\footnote{\
Greek indices vary from 0 to 3, Latin ones from 1 to 3 (spatial 
indices). Semi-colon means covariant derivative using the connection associated with the 
``physical" (curved) space-time metric, the latter being denoted by $\Mat{\gamma} $. 
Indices are raised and lowered with the help of this metric, unless 
explicitly mentioned otherwise.
}
\be\label{Lorentz force ETG}
F^i = q\,c \left ( \frac{F^i_{\ \, 0 }}{\beta } +  F^i_{\ \, j }\,\frac{v^j}{c} \right) = q\,F^i_{\ \, \mu }\,\frac{\dd x^\mu}{\dd t_{\bf x}}.
\ee
Here $v^j\equiv \dd x^j/\dd t_{\bf x} \ (j=1,2,3)$ is the velocity of the particle, measured with the local time $t_{\bf x}$, with
\be\label{dt_x}
\frac{\dd t_{\bf x}}{\dd t} = \beta (t,{\bf x}),
\ee
where $t=x^0/c$ is the coordinate time in a coordinate system $(x^\mu)$ adapted to (or bound with) the preferred reference fluid $\mathcal{E}$ assumed by the theory, and such that the synchronization condition $\gamma _{0 j}=0$ is true \cite{A54}; ${\bf x}\equiv (x^i)$; and (in any such coordinates)
\be
\beta \equiv  \sqrt{\gamma _{00}} . 
\ee
Equation (\ref{Lorentz force ETG}) may be rewritten in space-vector form as
\be\label{Lorentz force ETG-vector}
{\bf F} = q \left ( {\bf E} +  {\bf v}\wedge \,{\bf B} \right),\quad \quad ({\bf a}\wedge {\bf b})^i \equiv e  ^i_{\ \, jk}\,a^j\,b^k,
\ee
where the electric and magnetic vector fields are the spatial vector fields with components 
\be\label{E and B}
E^i\equiv \frac{c\,F^i_{\ \,0}}{\beta }, \quad B^k\equiv -\frac{1}{2}e^{ijk} F_{ij}.
\ee
In Eqs. (\ref{Lorentz force ETG-vector}) and (\ref{E and B}), $e_{ ijk}$ is the usual antisymmetric spatial tensor, its indices being raised or lowered using the spatial metric $\Mat{g}$ in the preferred frame $\mathcal{E}$; in spatial coordinate systems whose natural basis is direct, we have 
\be
e_{ ijk}=\sqrt{g}\, \varepsilon_{ ijk},\qquad e^{ ijk}=\frac{1}{\sqrt{g}}\, \varepsilon_{ ijk}, 
\ee
with $\varepsilon _{ ijk}$ the signature of the permutation $(i\,j\,k)$ and
\be
g \equiv \mathrm{det}(g_{ij}). 
\ee
For a continuous charged medium, the electric charge density is defined as 
\be
\rho 
_\mathrm{el} \equiv \delta q/\delta V, 
\ee
where $\delta V$ is the volume element measured with the physical volume measure: 
\be\label{delta V}
\delta V = \sqrt{ g}\, \dd x^{1}\dd x^{2}\dd x^{3}.
\ee
The 4-current is 
\be
J^{\, \mu } \equiv \rho_\mathrm{el}\,\dd x^\mu /\dd t_{\mathbf{x}}. 
\ee
The Lorentz force density is written, in accordance with (\ref{Lorentz force ETG}), as
\be\label{Lorentz force density}
f^i \equiv \frac{\delta F^i}{\delta V} =F^i_{\ \, \mu }\,J^\mu.
\ee

Dynamics of a test particle is defined by an extension to curved spacetime of the special-relativistic form of Newton's second law \cite{A16}. This applies to the non-interacting particles that constitute a {\it dust}. The following dynamical equation for a continuous medium with velocity field ${\bf v}$, subjected to a non-gravitational external force density field ${\bf f}$, has thus been {\it derived} for a dust from Newton's second law, and has been assumed to stay valid for a general continuum \cite{A54}:
\be\label{Eq T-f}
T_\mathrm{medium \ \,;\nu}^{0 \nu} =b^0(\Mat{T}_\mathrm{medium})+\frac{{\bf f.v}}{c\beta},
\qquad T_\mathrm{medium\ \,;\nu}^{i \nu} =b^i(\Mat{T}_\mathrm{medium})+ f^i,
\ee
where $\Mat{T}_\mathrm{medium}$ is the energy-momentum tensor of the continuous medium and
\be\label{b^mu}
b^0(\Mat{T}) \equiv \frac{1}{2}\,\gamma^{00}\,g_{ij,0}\,T^{ij},
\quad b^i(\Mat{T}) \equiv \frac{1}{2}\,g^{ij}g_{jk,0}\,T^{0k}.
\ee
For a {\it charged continuum,} $f^i$ is given by Eq. (\ref{Lorentz force density}), and the energy-momentum tensor $\Mat{T}_\mathrm{charged\ medium}$ has to be substituted for $\Mat{T}_\mathrm{medium}$. Then Eq. (\ref{Eq T-f}) for this medium can be rewritten as:
\be\label{Eq T-charged}
T^{\mu  \nu}_{\mathrm{charged\ medium}\ \ ;\nu} =b^\mu (\Mat{T}_\mathrm{charged\ medium})+F^\mu_{\ \, \nu }\,J^\nu.
\ee
It seems natural, almost obvious, to assume (i) that the \textit{total} energy-momentum is the sum 
\be\label{T=Tcharges+Tfield}
\Mat{T} = \Mat{T}_\mathrm{charged\ medium} + \Mat{T}_\mathrm{field}, 
\ee
where $\Mat{T}_\mathrm{field}$ is the energy-momentum tensor of the electromagnetic field \cite{L&L, Fock1964}:
\be\label{T em}
T_\mathrm{field}^{\, \mu \nu } \equiv \left (- F^\mu_{\ \ \lambda } F^{\, 
\nu \lambda } + \frac{1}{4}\gamma^{\, \mu \nu } F_{\lambda 
\rho } F^{\lambda \rho \, } \right)/\mu_0 , 
\ee
or
\be\label{T em-dev}
\Mat{T}_\mathrm{field} \equiv (T^\mu _{_\mathrm{field}\ \ \nu} ) = \left[\Mat{F} ^2 -\frac{1}{4} (\mathrm{tr}\,\Mat{F}^2)\Mat{I}\right]/\mu _0, \quad \Mat{F}\equiv (F^\mu _{\ \ \nu });
\ee
and (ii) that the total tensor $\Mat{T}$ obeys the general equation (\ref{Eq T-f}) for continuum 
dynamics, without any non-gravitational external force:
\be\label{Eq T}
T^{\mu  \nu}_{\ \ ;\nu} =b^\mu (\Mat{T}).
\ee
These two assumptions have indeed been made in Ref. \cite{A54}. Combining Eqs. (\ref{Eq T-charged}) and (\ref{Eq T}) using Eqs. (\ref{T=Tcharges+Tfield}) and (\ref{T em}), one derives
\be\label{Eq T-field-3}
F^\mu_{\ \ \lambda }\,F^{\lambda \nu }_{\ \ \,;\nu }= \mu_0 \left[ b^\mu \left (\Mat{T}_\mathrm{field} \right)-F^\mu_{\ \ \lambda }\,J^\lambda \right ],
\ee
where $b^\mu \left (\Mat{T}_\mathrm{field} \right)$ is given by Eqs. (\ref{b^mu}) and (\ref{T em}). Under the two assumptions right above, this is the second group of the gravitationally-modified Maxwell equations in the investigated theory --- at least 
for the generic case where the field tensor $\Mat{F}$ is invertible (det 
$\Mat{F}\equiv $ det ($F^\mu _{\ \ \nu}) \ne 0$). Indeed, if the matrix ($F^\mu _{\ \ \nu })$ is invertible, (\ref{Eq T-field-3}) is equivalent to
\be\label{Maxwell ETG}
F^{\mu \nu } _{\ \ \,;\nu  }=\mu_0 \left (G^\mu  _{\ \ \nu }\,b^\nu (\Mat{T}_\mathrm{field}) - J^\mu \right),\quad (G^\mu  _{\ \ \nu })\equiv (F^\mu  _{\ \ \nu })^{-1}.
\ee
(Note that $\Mat{G}$, like $\Mat{F}$, is an antisymmetric tensor, $G_{\nu \mu } = - G_{\mu \nu }$.)

\section{Charge balance: exact equations}\label{Non-conserv-exact}

Due to the antisymmetry of $\Mat{F}$, we have $F^{\mu \nu } _{\ \,;\nu;\mu   }=0$, and 
we get from Eq. (\ref{Maxwell ETG}):
\be\label{Charge rate ETG}
\hat{\rho  } \equiv \left ( J^\mu \right )_{; \mu}= \left( G^\mu  _{\ \,\nu }\,b^\nu (\Mat{T}_\mathrm{field}) \right )_{; \mu}.
\ee
Thus, $J^\mu _{; \mu} = 0$ --- which, as is well known, means exact charge conservation --- is not true in general, according to Eq. (\ref{Maxwell ETG}). To relate $\hat{\rho  } $ with the charge production or destruction in some ``substantial" domain $\Omega $ of the charged continuum, we note that, in coordinates adapted to (or bound with) this continuum $\mathcal{C}$, we have
\be
\sqrt{g_\mathcal{C}} = \sqrt{-\gamma }\, U^0,
\ee
with $g_\mathcal{C} \equiv \mathrm{det}(g_{ij\,\mathcal{C}})$, $\Mat{g}_\mathcal{C}$ being the spatial metric in the reference fluid $\mathcal{C}$, $\gamma \equiv \mathrm{det}(\gamma _{\mu \nu })$, and $U^\mu \equiv \dd x^\mu /\dd s$ the four-velocity field of the continuum. (We used the well-known general relation \cite{L&L}
\be\label{gamma}
\gamma =-\gamma _{0 0} g.)
\ee 
Therefore, noting $\rho^\star _\mathrm{el} \equiv \delta q/\delta V_\mathcal{C}$ the proper charge density, we have
\be
\frac{\dd}{\dd t} \left( \int_\Omega \delta q \right) = \int_\Omega \frac{\partial (\rho^\star _\mathrm{el} \sqrt{g_\mathcal{C}})}{\partial t}\,\dd^3 x = c \int_\Omega (\rho^\star _\mathrm{el}\,U^\mu)_{;\mu}\, \sqrt{-\gamma }\, \dd^3 x.
\ee
The two equalities are true (only) in coordinates adapted to $\mathcal{C}$. However, the rightmost integral is invariant under {\it any} change of the spatial coordinates, i.e., $x'^i=\psi^i(x^0,(x^j))$. Also the time coordinate $x^0$, with $t\equiv x^0/c$, is arbitrary. Thus we have in any coordinates $x^\mu $:
\be\label{Qpoint}
\frac{\dd}{\dd t} \left( \int_\Omega \delta q \right) = \int_\Omega \hat{\rho}\, \sqrt{-\gamma }\, \dd^3 x.
\ee
Of course the domain $\Omega $ as well as its boundary depend on $t$ in general spatial coordinates $x^i$.

\section{Weak field approximation}\label{PN}

We assume that the system of interest, $\mathrm{S}$, e.g. the solar system or even only the Earth, has a weak and slowly varying {\it  gravitational field}. To take benefit of this, we use a post-Newtonian (PN) approximation scheme that is in accordance with the general principles of asymptotic analysis. See Ref. \cite{A35}, Sect. V, and references therein. To do this, we conceptually associate with $\mathrm{S}$ a family $(\mathrm{S}_\lambda )$ of gravitating systems, depending smoothly on a parameter $\lambda $, so that the fields can be assumed to have asymptotic expansions as $\lambda \rightarrow 0$. In fact we need only low-order Taylor expansions at $\lambda =0$, which necessarily exist if the fields indeed depend smoothly on $\lambda $. Our assumption of a weak field then amounts to say that $\mathrm{S}$ corresponds to a small value of $\lambda $, say $\lambda _0 \ll 1$, so $\mathrm{S}=\mathrm{S}_{\lambda _0}$. Thus we can use the asymptotic expansions to approximate the values of the fields in that system by neglecting the remainder term. Each system is made of perfect fluids. The orders in $\lambda $ of the corresponding fields: pressure, density, velocity, and the scalar gravitational field $V\equiv -c^2 \mathrm{Log} \beta $ with $\beta \equiv \sqrt{\gamma _{00}}$, are the same as in a remarkable Newtonian similarity transformation \cite{FutaSchutz,A23}. It follows that by adopting $[\mathrm{M}]_\lambda = \lambda[\mathrm{M}]$ and $[\mathrm{T}]_\lambda = [\mathrm{T}]/\sqrt{\lambda}$  as the new units for the system $\mathrm{S}_\lambda$ (where $[\mathrm{M}]$ and $[\mathrm{T}]$ are the starting units of mass and time), all these fields become order $\lambda^0$, and the small parameter $\lambda$ is proportional to $1/c^2$; indeed $\lambda=(c_0/c_\lambda )^2$, where $c_0$ is the velocity of light in the starting time unit $[\mathrm{T}]$, and $c$ or $c_\lambda $ is with the time unit $[\mathrm{T}]_\lambda$. It thus becomes easy to derive asymptotic expansions. In particular, we have \cite{A35}:
\be\label{1-U/c2}
\beta \equiv \sqrt{\gamma _{00}} = 1 - U/c^2 + O(c^{-4}).
\ee
where $U$ is the zero-order term in the expansion of $V$. Moreover, the expansions are written in these $\lambda $-dependent units, in particular with the time $T$ (the preferred time of the theory \cite{A54,A35}) being counted with the unit $[\mathrm{T}]_\lambda$ for the system $\mathrm{S}_\lambda $, thus (\ref{1-U/c2}) should be written more precisely:
\be\label{1-U/c2-precise}
\beta^{(\lambda )}(T, {\bf x}) = 1 - U(T,{\bf x})/c_\lambda ^2 + O(c_\lambda ^{-4}).
\ee
Note that $U(T,{\bf x})$ is a coefficient in this asymptotic expansion and thus does not depend on $\lambda $. One may differentiate the PN expansions like (\ref{1-U/c2-precise}) with respect to the corresponding time variable, here $T\equiv x^0/c$ (as well as with respect to the spatial coordinates $x^i$)
\cite{A35}. This automatically accounts for the slow time variation. By doing this in the assumed wave equation for $V$, we obtain that $U$ in Eq. (\ref{1-U/c2}) or (\ref{1-U/c2-precise}) obeys  the Poisson equation with, on its r.h.s., the zero-order term in the expansion of the active mass density, hence $U$ is the Newtonian potential \cite{A35}. We obtain also that for the spatial metric, assumed in the theory to have the form
\be\label{Space metric-v2}
\Mat{g}=\beta ^{-2} \Mat{g}^0
\ee
(with $\Mat{g}^0$ an invariable Euclidean metric), we have:
\be\label{dg/dT}
\frac{\partial g_{ij}}{\partial T} = 2 c^{-2} \partial _T U \delta _{ij} +O(c^{-4})
\ee
(adopting Cartesian coordinates for the Euclidean metric $\Mat{g}^0$ until the end of Sect. \ref{Hertz}).\\

Regarding now the {\it electromagnetic field:} we assume that the field tensor $\Mat{F}$, as well as the current 4-vector ${\bf J}$, also depend smoothly on $\lambda $, hence they too admit low-order Taylor expansions at $\lambda =0$ --- but {\it a priori} we cannot say anything about the order of the main terms, hence we simply set (in the $\lambda $-dependent time and mass units, see above):
\be\label{expans-F}
\Mat{F} = c^n \left(\overset{0}{\Mat{F}} + c^{-2}\,\overset{1}{\Mat{F}} +O(c^{-4})\right) 
\ee 
and
\be\label{expans-J}
{\bf J} = c^m \left(\overset{0}{{\bf J}} + c^{-2}\,\overset{1}{{\bf J}} +O(c^{-4})\right),
\ee 
 for some integers $n$ and $m$ (positive, negative, or zero). Moreover, in contrast to the case with the gravitational field, we do not assume that $\Mat{F}$ is slowly varying (nor weak, as witnessed by the arbitrariness of $n$). This means that the expansions (\ref{expans-F})--(\ref{expans-J}) are not PN ones but post-Minkowskian (PM) expansions. We developed an asymptotic PM approximation scheme \cite{A34} using the same method as the one we developed for the PN approximation \cite{A35,A23}, in particular the time $T$ is counted with the unit $[\mathrm{T}]_\lambda$ for the system $\mathrm{S}_\lambda $ as is the case with PN expansions, but the time variable now is $x^0 =cT$. \footnote{\ 
Let $\mathrm{S}$ be a self-gravitating system of perfect fluids producing a weak gravitational field. The PM approximation associates with $\mathrm{S}$ a family $(\mathrm{S}_\lambda )$ of systems for which the velocity field ${\bf u}^{(\lambda)}$ is order zero in $\lambda $, not order $\sqrt{\lambda }$, hence it does not fit with the Newtonian limit. Nevertheless, it can be applied to a given such physical system $\mathrm{S}$ even if the fields really are slowly time-varying: the latter assumption just is not used. See the discussion in Ref. \cite{A34}, \S 4.1. However, in the case of an e.m. wave, we expect that it really is not varying slowly, of course.
}
One may differentiate PM expansions with respect to that new time variable. Therefore, in the modified Maxwell equation (\ref{Eq T-field-3}), the term $F^{\lambda \nu }_{\ \ \,;\nu }$ is of order $c^n$ as is also the term $F^\mu_{\ \ \lambda }$. On the r.h.s., we note that $\mu_0$ has dimension $MLQ^{-2}$, hence with the mass unit $[\mathrm{M}]_\lambda = \lambda[\mathrm{M}]$ we have $\mu _0=\mu _{0 0}\lambda ^{-1}$, with $\mu _{0 0}$ the value of $\mu _0$ in the starting units. That is, 
\be\label{expans-mu_o}
\mu _0=\mu _{0 0}\,c^2 
\ee
if we take $c_0=1$ for the simplicity of writing. Thus the r.h.s. of (\ref{Eq T-field-3}) is of order $c^{n+m+2}$ while the l.h.s. is of order $c^{2n}$, hence we must have
\be
m+n+2 = 2n,\qquad \mathrm{or}\quad m=n-2.
\ee
From (\ref{b^mu}), (\ref{dg/dT}), and (\ref{expans-F}), we get without difficulty
\be\label{expans-b^mu}
\mu _0 b_\mu(\Mat{T}_\mathrm{field}) = c^{2n-3}\,\partial _T U\times \left[ \overset{0}{T}\,^{j j}\ (\mu=0)\  \,or\   \,-\overset{0}{T}\,^{0 i}\ (\mu=i)   \right ] +O(c^{2n-5}),
\ee
where, in accordance with (\ref{T em-dev}), the matrix of the mixed components $\overset{0}{\Mat{T}}\equiv (\overset{0}{T}\,^\lambda _{\ \, \nu})$ is 
\be\label{T^0}
\overset{0}{\Mat{T}} = \overset{0}{\Mat{F}}\, ^2 -\frac{1}{4} (\mathrm{tr}\,\overset{0}{\Mat{F}}\, ^2)\Mat{I}.
\ee
By entering the expansions (\ref{expans-F})--(\ref{expans-J}) of $\Mat{F}$ and ${\bf J}$ into  (\ref{Eq T-field-3}) using this, we get the lowest-order term in the weak-field expansion of (\ref{Eq T-field-3}) as
\be\label{expans-Max2}
\overset{0}{F}\,^\mu_{\ \ \lambda }\,\overset{0}{F}\,^{\lambda \nu }_{\ \ \,,\nu }= -\mu_{00}\, \overset{0}{F}\,^\mu_{\ \ \lambda }\,\overset{0}{J}\,^\lambda.
\ee
(We used also the identity $F^{\lambda \nu }_{\ \ \,;\nu }=(\sqrt{-\gamma }F^{\lambda \nu }_{\ \ \,,\nu })/\sqrt{-\gamma }$ \cite{L&L} and the fact that, from (\ref{gamma}) and (\ref{Space metric-v2}), we have 
\be\label{sqrt gamma}
\sqrt{-\gamma }=1+O(c^{-2}).) 
\ee
Equation (\ref{expans-Max2}) is an exact equation for the expansion coefficient fields $\overset{0}{\Mat{F}}$ and $\overset{0}{{\bf J}}$, as is usual with an asymptotic expansion \cite{A23}. Hence, when the field matrix $\overset{0}{\Mat{F}}\equiv (\overset{0}{F}\,^\lambda _{\ \, \nu})$ is invertible, that field is an exact solution of the flat-spacetime Maxwell equation:
\be\label{flat-Maxwell-wrong-dim}
\overset{0}{F}\,^{\lambda \nu }_{\ \ \,,\nu }= -\mu_{00}\,\overset{0}{J}\,^\lambda.
\ee
On the other hand, using (\ref{expans-F}), (\ref{expans-b^mu}), and (\ref{sqrt gamma}) in (\ref{Charge rate ETG}) gives 
us 
\be\label{expans-rho-hat}
\hat{\rho } = c^{n-5} \mu_{00}^{\ \  -1} \left[ \left( \overset{0}{G}\,^{\mu  0}\, \overset{0}{T}\,^{j j}- \overset{0}{G}\,^{\mu i}\, \overset{0}{T}\,^{0 i}\right) \partial _T U \right]_{,\mu }+O(c^{n-7}), 
\ee
where $\overset{0}{\Mat{G}}\equiv (\overset{0}{G}\,^\mu _{\ \,\nu }) \equiv \overset{0}{\Mat{F}}\,^{-1}$. Due to (\ref{expans-F})-(\ref{expans-J}), $\overset{0}{\Mat{F}}$, $\overset{0}{\Mat{G}}$, $\overset{0}{\Mat{T}}$ and $\overset{0}{{\bf J}}$ do not have the physical dimensions of the corresponding fields $\Mat{F}, \Mat{G},\,etc.$
. Therefore, in accordance with (\ref{expans-F})-(\ref{expans-J}), let us define
\be\label{FJ1 from FJ0}
\Mat{F}_1 \equiv c^{n}\overset{0}{\Mat{F}}\quad \mathrm{and}\quad {\bf J}_1 \equiv c^{m}\overset{0}{{\bf J}}.
\ee
In view of (\ref{expans-mu_o}) and (\ref{flat-Maxwell-wrong-dim}), $\Mat{F}_1$ and  ${\bf J}_1$ are solutions of the flat-spacetime Maxwell equation with the correct dimensions in the SI units: 
\be\label{flat-Maxwell}
(F_1)\,^{\lambda \nu }_{\ \ \,,\nu }= -\mu_{0}\,(J_1)\,^\lambda.
\ee
Obviously, the matrix $\Mat{F}_1$ is invertible when $\overset{0}{\Mat{F}}$ is, and its inverse is $\Mat{G}_1\equiv c^{-n}\overset{0}{\Mat{G}}$. Also, from (\ref{expans-mu_o}) and (\ref{T^0}), the e.m. T-tensor $\Mat{T}_1$ associated with $\Mat{F}_1$ by (\ref{T em-dev}) is 
\be
\Mat{T}_1 = \mu_{0 0}^{-1} c^{2n-2} \overset{0}{\Mat{T}}.
\ee
With this, we can rewrite (\ref{expans-rho-hat}) as 
\be\label{expans-rho-hat-2}
\hat{\rho } = c^{-3}  \left[ \left( G_1\,^{\mu  0}\, T_1\,^{j j}- G_1\,^{\mu i}\, T_1\,^{0 i}\right) \partial _T U \right]_{,\mu }+O(c^{-5}).
\ee
Note that this is {\it independent} of the integers $n$ and $m$ in Eqs. (\ref{expans-F})-(\ref{expans-J}). \\

Also note that, substituting (\ref{expans-F}) into the first group (\ref{Maxwell 1}), one finds that $\overset{0}{\Mat{F}}$ and hence $\Mat{F}_1$ are also exact solutions of (\ref{Maxwell 1}). To use the result (\ref{expans-rho-hat-2}) so as to assess the charge production that could be expected as a consequence of the modified second group (\ref{Eq T-field-3}), we have conversely to assume the following: Let $\Mat{F}_1$ and  ${\bf J}_1$ be solutions of the complete flat-spacetime Maxwell equations, i.e., the first group (\ref{Maxwell 1}) and the second group (\ref{flat-Maxwell}). (Note that the first group (\ref{Maxwell 1}) is the same independently of the presence or absence of a gravitational field.) In a given weak gravitational field, to any such pair $(\Mat{F}_1,{\bf J}_1)$ there corresponds a unique solution $(\Mat{F},{\bf J})$ of the first group (\ref{Maxwell 1}) and the {\it modified} second group (\ref{Eq T-field-3}), such that $(\Mat{F}_1,{\bf J}_1)$ be the first terms in the PM expansion of $(\Mat{F},{\bf J})$, i.e., we have (\ref{FJ1 from FJ0}). This is expectable from perturbative arguments.

\section{Explicit expressions}\label{Explicit}

To use (\ref{expans-rho-hat-2}) we need explicit expressions,  in terms of the components $E^i$ and $B^i$ of the electric and magnetic fields, for the field matrix $\Mat{F}\equiv (F^\lambda _{\ \,\nu })$, the corresponding T-tensor $\Mat{T}$, the inverse matrix $\Mat{G}\equiv (G^\lambda _{\ \,\nu })\equiv \Mat{F}^{-1}$, and the 4-vector ${\bf e}$ with 
\footnote{\label{Covar e^mu}\
Like in Sect. \ref{PN}, the equations in this section are valid in spatial coordinates that are adapted to the frame $\mathcal{E}$ and, more specifically, are Cartesian for the Euclidean spatial metric $\Mat{g}^0$; and with the time coordinate being $x^0=cT$, the preferred time coordinate of the theory. However, Eqs. (\ref{F-matrix})--(\ref{T^0 i}) are valid more generally in any Cartesian coordinates for the flat spacetime metric $\Mat{\gamma }^0$ that is built \cite{A35} with $\Mat{g}^0$ and $T$. We use $E_i\equiv E^i$ and  $B_i\equiv B^i$, which means lowering the index $i$ {\it for ${\bf E}$ and ${\bf B}$} with the {\it spatial,} Euclidean part of metric $\Mat{\gamma }^0$ in such Cartesian coordinates.
}
\be\label{emu}
e^\mu \equiv G\,^{\mu  0}\, T\,^{j j}- G\,^{\mu i}\, T\,^{0 i}.
\ee
(We omit the index $1$ used in (\ref{flat-Maxwell})-(\ref{expans-rho-hat-2}).) We have 
\be\label{F-matrix}
\Mat{F} \equiv (F^\mu_{\ \ \nu}) = \left(\begin{array}{cccc} 0 & E_1/c & E_2/c & E_3/c\\ E_1/c & 0 & B_3 & - B_2\\ E_2/c & - B_3 & 0 & B_1\\ E_3/c & B_2 & - B_1 & 0 \end{array}\right).
\ee
When $\mathrm{det}\Mat{F} \ne 0$, the inverse matrix $\Mat{G} = \Mat{F}^{-1}$ has a known expression for a general $4\times4$ matrix:
\be\label{Inv-4x4}
\Mat{G} = \frac{1}{\mathrm{det}\Mat{F}} \left[\frac{1}{6}\left((\mathrm{tr}\, \Mat{F})^3-3\mathrm{tr}\, \Mat{F}\,\mathrm{tr}\, \Mat{F}^2+2\mathrm{tr}\, \Mat{F}^3 \right )\Mat{I} -\frac{1}{2}\left( (\mathrm{tr}\, \Mat{F})^2-\mathrm{tr}\, \Mat{F}^2\right)+ \Mat{F}^2\,\mathrm{tr}\, \Mat{F}-\Mat{F}^3\right].
\ee
Here, $\Mat{F}$ has the following block structure:
\be\label{Fblock}
\Mat{F}=\left (\begin{array}{cc} 0 \quad l \\ l^T \quad a \end{array} \right),
\ee
where $l$ is a $1\times 3$ matrix (row) and $a$ is a $3\times 3$ antisymmetric matrix. Clearly, $\mathrm{tr}\, \Mat{F}=0$. After computing $\Mat{F}^3$ in terms of this block structure, one finds that $\Mat{F}^3$ has all diagonal terms zero, so $\mathrm{tr}\, \Mat{F}^3=0$, too. Using this in (\ref{F-matrix}) and (\ref{Inv-4x4}), with Matlab Symbolic Toolbox we obtain for $\Mat{G}' \equiv (G^{\mu \nu })$ the following matrix: 
\be\label{G-matrix}
 \left(\begin{array}{cccc} 0 & -\frac{B_1\, c}{B_1\, E_1 + B_2\, E_2 + B_3\, E_3} & -\frac{B_2\, c}{B_1\, E_1 + B_2\, E_2 + B_3\, E_3} & -\frac{B_3\, c}{B_1\, E_1 + B_2\, E_2 + B_3\, E_3}\\ \frac{B_1\, c}{B_1\, E_1 + B_2\, E_2 + B_3\, E_3} & 0 & \frac{E_3}{B_1\, E_1 + B_2\, E_2 + B_3\, E_3} & -\frac{E_2}{B_1\, E_1 + B_2\, E_2 + B_3\, E_3}\\ \frac{B_2\, c}{B_1\, E_1 + B_2\, E_2 + B_3\, E_3} & -\frac{E_3}{B_1\, E_1 + B_2\, E_2 + B_3\, E_3} & 0 & \frac{E_1}{B_1\, E_1 + B_2\, E_2 + B_3\, E_3}\\ \frac{B_3\, c}{B_1\, E_1 + B_2\, E_2 + B_3\, E_3} & \frac{E_2}{B_1\, E_1 + B_2\, E_2 + B_3\, E_3} & -\frac{E_1}{B_1\, E_1 + B_2\, E_2 + B_3\, E_3} & 0 \end{array}\right).
\ee
(We checked that the product $\Mat{F}.\Mat{G}$ is exactly the identity matrix $\Mat{I}$.) For the combinations (\ref{expans-rho-hat-2}) of the T-tensor (\ref{T em-dev}), we get from (\ref{F-matrix}) well-known results:
\be\label{W_field}
T^{jj} =
\frac{{B_1}^2\, c^2 + {B_2}^2\, c^2 + {B_3}^2\, c^2 + {E_1}^2 + {E_2}^2 + {E_3}^2}{2\, c^2\, \mu_0} \equiv W_\mathrm{field},
\ee
\be\label{T^0 i}
T^{0\,i}=\left(\begin{array}{ccc} -\frac{B_2\, E_3 - B_3\, E_2}{c\, \mu_0} & \frac{B_1\, E_3 - B_3\, E_1}{c\, \mu_0} & -\frac{B_1\, E_2 - B_2\, E_1}{c\, \mu_0} \end{array}\right).
\ee
Putting this in (\ref{emu}) using (\ref{G-matrix}) gives us
\be\label{e^mu}
e^\mu =\left(\begin{array}{c} 0\\ \frac{{B_1}^3\, c^2 + B_1\, {B_2}^2\, c^2 + B_1\, {B_3}^2\, c^2 + B_1\, {E_1}^2 - B_1\, {E_2}^2 - B_1\, {E_3}^2 + 2\, B_2\, E_1\, E_2 + 2\, B_3\, E_1\, E_3}{2\, c\, \mu_0\, \left(B_1\, E_1 + B_2\, E_2 + B_3\, E_3\right)}\\ \frac{{B_1}^2\, B_2\, c^2 + 2\, B_1\, E_1\, E_2 + {B_2}^3\, c^2 + B_2\, {B_3}^2\, c^2 - B_2\, {E_1}^2 + B_2\, {E_2}^2 - B_2\, {E_3}^2 + 2\, B_3\, E_2\, E_3}{2\, c\, \mu_0\, \left(B_1\, E_1 + B_2\, E_2 + B_3\, E_3\right)}\\ \frac{{B_1}^2\, B_3\, c^2 + 2\, B_1\, E_1\, E_3 + {B_2}^2\, B_3\, c^2 + 2\, B_2\, E_2\, E_3 + {B_3}^3\, c^2 - B_3\, {E_1}^2 - B_3\, {E_2}^2 + B_3\, {E_3}^2}{2\, c\, \mu_0\, \left(B_1\, E_1 + B_2\, E_2 + B_3\, E_3\right)} \end{array}\right).
\ee
Note that $e^0=0$ comes simply from the fact that ${\bf B.}({\bf E}\wedge {\bf B})=0$. Thus (\ref{expans-rho-hat-2}) rewrites as 
\be\label{expans-rho-hat-3}
\hat{\rho } = c^{-3}  \left( e^i \partial _T U \right)_{,i }+O(c^{-5}).
\ee

To assess $\partial_T U$ and $\partial _T(\nabla U)$, we note that these time derivatives must be evaluated in the preferred reference frame $\mathcal{E}$ assumed by the theory, and that the system of interest producing the e.m. field is expected to have a motion through that ``ether", with a velocity field ${\bf v}$ whose modulus may be in the range 10-1000 km/s. In that situation, the main contribution to $\partial_T U$ is that which is due to the mere translation motion of the relevant astronomical body --- say the Earth --- through $\mathcal{E}$. Indeed, from the well-known integral giving $U$ in terms of the Newtonian (i.e., zero-order PN) mass-energy density $\rho $, one gets exactly 
\be
\dd U/\dd T \equiv \partial_T U +{\bf v.}\nabla U = 0
\ee 
if the system producing $U$ has a rigid motion. In particular, it is exact for the self potential of a body whose motion is rigid. This is true for the Earth to a very good approximation, moreover the potential due to the Sun is nearly constant also on Earth; the most important departure from $\dd U/\dd T=0$ should come from the Moon. For a rigidly rotating {\it spherical} body, we have moreover
\be
{\bf v.}\nabla U={\bf V}.\nabla U, 
\ee
with ${\bf V}\equiv \dot{{\bf a}} $, ${\bf a}(T)$ being the center of the body. Thus to a good approximation (except for an unexpectedly small velocity $V$), largely sufficient to get an order-of-magnitude estimate, we may consider the Earth as an isolated, rigidly moving, spherically symmetric body. For such a body we have:
\be\label{d_T U}
\partial_T U = -{\bf V.}\nabla U = \frac{GM(r)}{r^2}{\bf V.e}_r, \quad r\equiv \abs{{\bf x}-{\bf a}(T)}, \quad {\bf e}_r \equiv ({\bf x}-{\bf a}(T))/r,
\ee
where
\be
M(r) \equiv 4 \pi \int_0 ^r u^2 \rho (u)\,\dd u.
\ee
Let us compute $\partial _T(\nabla U)$, with $\nabla U = -\frac{GM(r)}{r^2}{\bf e}_r$. To do that we may assume that ${\bf a}(T)={\bf V}T$. We then get 
\be
\partial_T \nabla U = -G \left \{\left [ \left({\bf V}^2 T-{\bf V.x}\right ) \left(4\pi  \frac{\rho(r)}{r}- \frac{3M(r)}{r^4} \right )\right ]\,{\bf e}_r - \frac{M(r){\bf V}}{r^3} \right \}.
\ee
On the r.h.s., the two terms inside the large parentheses have the same order of magnitude as the last term inside the braces. Moreover those two terms cancel one another if the rotating spherical body is homogeneous. In that case we thus have:
\be\label{d_T gradU}
\partial_T \nabla U = \frac{GM(r)}{r^3}{\bf V}.
\ee
We shall use this approximation to get an order-of-magnitude estimate. On the Earth's surface, (\ref{d_T U}) gives  
\be
\partial_T U\simeq gV_r \lessapprox 10V\simeq 10^5 \ (\mathrm{MKSA})\ \mathrm{for}\ V=10\, \mathrm{km/s}\ (10^4\,\mathrm{MKSA}),
\ee 
and (\ref{d_T gradU}) gives 
\be\label{d_T gradU-num}
\partial_T \nabla U\simeq g{\bf V}/R,\quad \abs{\partial_T \nabla U} \simeq 1.5\times 10^{-2}\ (\mathrm{MKSA})\ \mathrm{for}\  V=10\,\mathrm{km/s}.
\ee

\section{Case of a plane wave}\label{Plane}

Let us consider a general plane e.m. wave, whose propagation direction may be assumed parallel to the Cartesian basis vector ${\bf i}\equiv \partial _1$:
\be\label{Plane wave}
E^1 = 0, \quad {\bf E}={\bf E}(x^1), \quad B^1=0, \quad {\bf B}={\bf B}(x^1),\quad c{\bf B}= {\bf i}\wedge {\bf E}.
\ee
This is a solution of the flat Maxwell equations {\it in vacuo}. The electric and magnetic fields are orthogonal for such a wave. For the field matrix $\Mat{F}\equiv (F^\mu_{\ \ \nu})$, most generally, the condition $\mathrm{det}\,\Mat{F}\ne 0$ is equivalent to ${\bf E.B}\equiv \Mat{g}({\bf E},{\bf B})\ne 0$ \cite{A54}. It follows that $\Mat{F}$  is not invertible for a plane wave. But the e.m. field 
\be\label{EBtilde}
(\tilde{{\bf E}}, \tilde{{\bf B}})\equiv ({\bf E}+{\bf E}', {\bf B}+{\bf B}'),
\ee
with $({\bf E}', {\bf B}')$ any constant e.m. field, is still a vacuum solution of the flat Maxwell equations depending only on $x^1$ --- in fact, it is still a ``plane wave" to the same extent as (\ref{Plane wave}). For that field, generically, $\tilde{\Mat{F}}$ is invertible. In that case, $e^i$ [Eq. (\ref{e^mu})] is well defined and has $e^i_{,i}=0$, because (imposing $E'^1 = B'^1 = 0$) $e^1=0$ and $e^i=e^i(x^1)$. Neglecting the term $c^{-3}e^i(\partial _TU)_{,i}$ in view of (\ref{d_T gradU-num}) and the extreme smallness of the $c^{-3}$ factor, we thus get from (\ref{expans-rho-hat-3}) for the field (\ref{EBtilde}):
 \be
\hat{\rho}=0\qquad (\mathrm{Plane\ wave,}\ c^{-3}e^i(\partial _TU)_{,i}\ \mathrm{neglected}).
 \ee
Since this is independent of the constant e.m. field $({\bf E}', {\bf B}')$, it remains true at the limit of a ``pure" plane e.m. wave (\ref{Plane wave}), for which $c{\bf B}={\bf i}\wedge {\bf E}$. However, the value of the neglected term does depend on the constant e.m. field $({\bf E}', {\bf B}')$, and can be large if that field is very strong.

\section{Case of a group of Hertzian dipoles}\label{Hertz}

Hertz's famous oscillating dipole is the electric charge {\it distribution}
\be\label{rho_Hertz}
\rho_\mathrm{el} = T_{{\bf d}, {\bf b},\omega } \equiv - e^{-i\omega t} \, {\bf d.} \nabla \delta _{\bf b}
\ee
(more exactly, $\rho_\mathrm{el}$ is the real part of the r.h.s.). Here: ${\bf b} $ is the dipole's 
position; ${\bf d}$ is the dipole vector, and $\delta _{\bf b}$ is Dirac's measure at 
${\bf b} $. The associated 3-current is (the real part of the r.h.s. in):
\be\label{j_Hertz}
{\bf j} = - i\omega {\bf d}\, e^{-i\omega t} \,  \delta _{\bf b},
\ee
and the corresponding conserved 4-current is ${\bf J} \equiv (c\rho_\mathrm{el},{\bf j})$. The following electric and magnetic fields (or the associated field matrix (\ref{F-matrix})) provide an exact solution of the flat Maxwell equations in the distributional sense, for that current ${\bf J}$:
\footnote{\
This solution can be easily found in the literature, e.g. Jackson \cite{Jackson1998}, McDonald \cite{McDonald2004}, but whether or not it is an exact solution does not appear clearly. We can prove that it is an exact solution; here is a summary. In the Lorenz gauge, the vector potential ${\bf A}$ obeys the wave equation: $\square {\bf A}=\mu _0 {\bf j}$ (\cite{Jackson1998}, Eq. (6.16)). For the case of a stationary source, the integral solution ${\bf A}$ simplifies to Eq. (9.3) in Ref. \cite{Jackson1998}. The potential (9.3) rewrites exactly as (9.11) \cite{Jackson1998}. When applied to the ``Dirac current" (\ref{j_Hertz}), we may easily show that (9.11) rewrites also {\it exactly} as (9.13). In turn, (9.13) leads to the solution (\ref{E_Hertz})--(\ref{B_Hertz}) \cite{Jackson1998}. That (9.3) applies to the singular current (\ref{j_Hertz}), can be checked from the general formula ${\bf A}=\mu _0 E\star {\bf S}$, with $E$ the elementary solution of the wave operator that has support in the half-space $x^0 \ge 0$ of $\mathbb{R}^4$ --- that is the distribution given by $\langle E, h \rangle = \int_{\mathbb{R}^3} h(\abs{{\bf x}},{\bf x})\, \dd {\bf x}/(4 \pi \abs{{\bf x}}),\quad h\in \mathcal{D}(\mathbb{R}^4)$ \cite{Guichardet1969}. That general formula for ${\bf A}$ is here applied with ${\bf S}$ the current (\ref{j_Hertz}): as a distribution acting on functions in $\mathcal{D}(\mathbb{R}^4)$, this current must be more precisely defined as ${\bf S}\equiv - i\omega {\bf d}\,U  \otimes \delta _{\bf b}$, with $U$ the distribution on $\mathbb{R}$ associated with the function $t\mapsto e^{-i\omega t}$.
}
\be\label{E_Hertz}
{\bf E} = \alpha \left \{ \frac{k^2}{r}\left ( {\bf d} - (\hat{{\bf r}}{\bf .d}) \hat{{\bf r}}\right) \cos \varphi + \left [3(\hat{{\bf r}}{\bf .d}) \hat{{\bf r}} - {\bf d}\right ] \left (\frac{\cos \varphi }{r^3}+ \frac{k\sin\varphi }{r^2} \right ) \right \},
\ee
\be\label{B_Hertz}
{\bf B} = \alpha '  k^2 (\hat{{\bf r}}\wedge {\bf d}) \left ( \frac{\cos \varphi }{r}-\frac{\sin\varphi }{kr^2} \right ), \quad k = \frac{\omega }{c}, \quad \varphi \equiv kr-\omega t .
\ee
Here, $\alpha \equiv \frac{1}{4 \pi \epsilon_0}= 9 \times 10^6,\ \ \alpha ' \equiv \frac{c}{4\pi}\simeq 2.39\times 10^7$ (MKSA). Thus, ${\bf E.B} =0$ since ${\bf E}$ is in the plane containing ${\bf d}$ and $\hat{{\bf r}}\equiv ({\bf x}-{\bf b})/\abs{{\bf x}-{\bf b}}$, while ${\bf B}\propto \hat{{\bf r}}\wedge {\bf d}$. However, we may consider a group of dipoles with different ${\bf b}$'s  and ${\bf d}$'s, and by adding the corresponding fields (\ref{E_Hertz})--(\ref{B_Hertz}) we get exact solutions of the flat Maxwell equations for which, generically, we have ${\bf E.B} \ne 0$.\\

We thus consider a group of dipoles that all are at rest in a common frame $\mathcal{E}_{\bf V}$ moving uniformly at ${\bf V}$ with respect to the preferred reference frame $\mathcal{E}$. 
To calculate $\hat{\rho}$ in the vicinity of those moving dipoles, we consider successively each cube $\mathrm{C}$ in a regular mesh of small cubes at rest in $\mathcal{E}_{\bf V}$. Each cube $\mathrm{C}$ is defined by $\abs{x^i-a^i}\leq h/2$, where $x^i\ (i=1,2,3)$ are Cartesian coordinates for the Euclidean metric $\Mat{g}'^0$ that is the spatial part of the flat Minkowski metric $\Mat{\gamma }^0$ (Note \ref{Covar e^mu}) in the inertial frame $\mathcal{E}_{\bf V}$. (We mean inertial in the Minkowski space $(\mathrm{V},\Mat{\gamma }^0)$, with $\mathrm{V}$ the spacetime manifold. The uniformity of ${\bf V}$ is meant in the same sense.) Neglecting the $O(c^{-5})$ remainder in (\ref{expans-rho-hat-3}), we have:
\be\label{int_hat_rho}
\int_{\mathrm{C}_\mathcal{E}(T)} \hat{\rho }(T,{\bf X})\, \dd ^3 X = c^{-3}  \int_{\mathrm{C}_\mathcal{E}(T)} \left( e^i \partial _T U \right)_{,i }\,\dd ^3 X=  c^{-3} \int _{\partial \mathrm{C}_\mathcal{E}(T)} \Mat{g}^0({\bf e},{\bf n})\partial _T U \, \, \dd S,
\ee 
where $\mathrm{C}_\mathcal{E}(T)$ is the cube $\mathrm{C}$, as it appears at the time $T$ in the frame $\mathcal{E}$, the latter being endowed with Cartesian coordinates $X^i\, (i=1,2,3)$ for the Euclidean metric $\Mat{g}^0$ (that is the spatial part of metric $\Mat{\gamma }^0$ in the inertial frame $\mathcal{E}$); we denote ${\bf X}\equiv (X^i)$; ${\bf n}$ is the external normal (for the metric $\Mat{g}^0$) to the boundary $\partial \mathrm{C}_\mathcal{E}(T)$, and $\dd S$ is the Euclidean surface element. The Cartesian coordinates $(X^i)=(X,Y,Z)$ and $(x^i)=(x,y,z)$ that we consider are such that the $Ox$ axis coincides with $OX$ and is parallel to ${\bf V}$, so we are in the conditions of a special Lorentz transformation $L_{\bf V}$ from $\mathcal{E}$ to $\mathcal{E}_{\bf V}$:
\be\label{L_V}
L_{\bf V}: t=\gamma _V(T-VX/c^2),\ x=\gamma_V (X-VT),\ y=Y,\ z=Z,
\ee
with $t$ the inertial time in the frame $\mathcal{E}_{\bf V}$ and $\gamma _V$ the Lorentz factor. \\

Let ${\bf x}\equiv (x,y,z)$ be a given point of the domain $\mathrm{C}$ that is at rest in the frame $\mathcal{E}_{\bf V}$. At the time $T$ of the preferred inertial frame, that point corresponds in the frame $\mathcal{E}$ to a spatial position ${\bf X}\equiv (X,Y,Z)$ such that
\be
(t,{\bf x}) = L_{\bf V}(T,{\bf X}).
\ee
The unknowns are ${\bf X}$ and the value $t$ of the time of $\mathcal{E}_{\bf V}$. (Here $Y=y$ and $Z=z$ from (\ref{L_V}), so the unknowns are $t$ and $X$.) Equation (\ref{L_V}) leads to
\be
t = \frac{T}{\gamma _V}- \frac{Vx}{c^2}.
\ee
Knowing $t$ and ${\bf x}$ we compute the sum of the fields (\ref{E_Hertz})--(\ref{B_Hertz}) from the different dipoles, which is thus got in the moving frame $\mathcal{E}_{\bf V}$, say ${\bf E}',{\bf B}'$. We then transform the field to the frame $\mathcal{E}$ by the inverse Lorentz transformation:
\bea
E_1 & = & E'_1,\ E_2=\gamma _V(E'_2+VB'_3),\ E_3=\gamma _V(E'_3-VB'_2),\\
B_1 & = & B'_1,\ B_2=\gamma _V(B'_2-VE'_3/c^2),\ B_3=\gamma _V(B'_3+VE'_2/c^2),
\eea
and hence can compute the components $e^i$ of ${\bf e}$ by Eq. (\ref{e^mu}). We can thus calculate the surface integral on the r.h.s. of (\ref{int_hat_rho}). From (\ref{int_hat_rho}), the value of the field at the center ${\bf a}$ of the cube $\mathrm{C}$ is then approximately
\footnote{\
We approximate the integral of ${\bf w.n}$ (with ${\bf w}$ a spatial vector field, here $w^j=e^j \partial_T U$) over an $i$ face of the rectangle parallelepiped $\mathrm{C}_\mathcal{E}(T)$ as $I\simeq \pm hh'(w^i(S_1)+...+w^i(S_4))/4$, where $h$ and $h'$ are the rectangular face's sides and $S_k\ (k=1,...,4)$ are the vertices of that face. 
This algorithm has been tested by applying it to simple vector fields, e.g. ${\bf w}(x,y,z)=(-xy,xy,z^2)$. As expected, we found that $\int _{\partial \mathrm{C}} w^i n_i\, \dd S/h^3$, thus calculated, approximates closely the value of $\mathrm{div}\,{\bf w}=w^j_{,j}$ at the centre ${\bf a}$ of a small cube $\mathrm{C}$ with side $h$.
}
\be\label{hat_rho}
\hat{\rho }(T,{\bf a}) \simeq c^{-3} \int _{\partial \mathrm{C}_\mathcal{E}(T)} e^i n_i\, \partial _T U \, \dd S/(h^3\gamma _V),
\ee 
since $h^3\gamma _V$ is the Euclidean volume of the rectangle parallelepiped $\mathrm{C}_\mathcal{E}(T)$, deduced from the small cube $\mathrm{C}$ by Lorentz contraction.\\

For three dipoles with $d=100\,$nC.m, $\nu  =100\,$MHz ($\lambda =3\,$m), situated at $\lessapprox \lambda $ from one another, we get fields with moduli $E \lessapprox $ a few $10^5 \,$V/m, $B\lessapprox 15\, $T (in the moving frame). With $V=10\,$km/s, $\hat{\rho }(T,{\bf x})$ (counted in electrons per period per cubic meter) then has peaks at $\approx \pm 2\times 10^8\,e/$m$^3/$period. (The peaks are very sharp and their values depend somewhat on the discretization. We also integrated in time and the values keep very high.) This seems untenable, even though the sign of the predicted charge production alternates in space. Therefore, the version proposed in Ref. \cite{A54} of the gravitationally-modified second group of Maxwell equations, Eq. (\ref{Eq T-field-3}) here, looks like being discarded.

\section{The reason for the problem}\label{Reason}

So it seems that Eq. (\ref{Eq T-field-3}) is not the right Maxwell second group of the theory. Why does this happen? As explained in Sect. \ref{Main}, Eq. (\ref{Eq T-field-3}) has been deduced \cite{A54}, under two assumptions, from the general dynamical equation for a continuum subjected to a non-gravitational external force density field $f^i$, Eq. (\ref{Eq T-f}) --- when the latter is applied to the case that the continuum is a charged one and the external force density $f^i$ is the Lorentz force density (\ref{Lorentz force density}). Those  \hypertarget{Ass-i}{two assumptions} 
are:\\

\noi (i) The total energy-momentum tensor is \hypertarget{Ass-ii}{the sum} $\Mat{T} = \Mat{T}_\mathrm{charged\ medium} + \Mat{T}_\mathrm{field}$.\\

\noi (ii) The total tensor $\Mat{T}$ obeys the general equation for continuum dynamics, without any non-gravitational force, Eq. (\ref{Eq T}).\\

As it has been noted in Ref. \cite{A54}, Eq. (\ref{Eq T-f}) applied to the charged continuum, plus Assumptions (i) and (ii), lead straightforwardly, in view of the linearity of the dependence $b^\mu=b^\mu(\Mat{T})$, to
\be\label{Eq T-f-field}
T_{\mathrm{field}\ \,;\nu}^{0 \nu} =b^0(\Mat{T}_\mathrm{field})-\frac{{\bf f.v}}{c\beta},
\qquad T_{\mathrm{field}\ \,;\nu}^{i \nu} =b^i(\Mat{T}_\mathrm{field})- f^i.
\ee
It is precisely this equation which, combined with Eq. (\ref{Lorentz force density}) giving the Lorentz force density and with the expression (\ref{T em}) of the energy-momentum tensor, leads to the second group (\ref{Eq T-field-3}) \cite{A54}. Now we observe that Eq. (\ref{Eq T-f-field})$_2$ has exactly the form of (\ref{Eq T-f})$_2$ as applied not to the charged medium but {\it to the electromagnetic ``field continuum" itself,} if and only if the density field of the non-gravitational external force on the field continuum is given by
\be\label{f_field}
f^i_\mathrm{field}= -f^i \equiv -f^i_\mathrm{charged\ medium}.
\ee
In addition, we observe that Eq. (\ref{Eq T-f-field})$_1$ has exactly the form of (\ref{Eq T-f})$_1$ as applied to 
the 
``field continuum", if and only if the velocity field of that continuum is well-defined and verifies the following relation:
\be\label{v_field}
{\bf f}_\mathrm{field}{\bf .v}_\mathrm{field}=-{\bf f .v}\equiv -{\bf f}_\mathrm{charged\ medium}{\bf .v} _\mathrm{charged\ medium},
\ee
that is, with Eq. (\ref{f_field}), if and only if
\be\label{f.(v_field-v_charges)=0}
{\bf f}_\mathrm{charged\ medium}{\bf .}\left({\bf v}_\mathrm{field} -{\bf v}_\mathrm{charged\ medium}\right)=0.
\ee
The force density ${\bf f}_\mathrm{field}$ should represent the reaction of the charged continuum to the Lorentz force exerted 
by the field continuum. Thus (\ref{f_field}) means the opposition of action and reaction. Poincar\'e has shown that this does 
not apply to ``matter" (the charged medium, e.g. a Hertz oscillator), but this was in the sense that matter emitting e.m. 
radiation does not conserve its momentum unless one counts also the momentum of the emitted e.m. field, i.e., the total momentum 
is in fact conserved  \cite{Poincare1900}. In the present case, the actio-reactio opposition suggested by Eq. (\ref{Eq 
T-f-field})$_2$ concerns the charged medium on one hand and the e.m. field on the other hand, thus it means that there is no net 
force on the combined medium: ``charged medium plus e.m. field" (except for gravitation). This is consistent with the 
conservation of the total momentum (in the absence of gravitation). \footnote{\
See \S III.C in Ref. \cite{A35} for a discussion of the total momentum in the presence of gravitation in the present theory.
}
As to the velocity of the field continuum: there is one situation for which it can be naturally deduced from its energy-momentum tensor, namely the case that the energy-momentum tensor is a tensor product, i.e. has the bilinear form
\be\label{T = tensor product}
T^{\mu \nu } = V^\mu  V^\nu . 
\ee
Equation (\ref{T = tensor product}) is true for a dust made of ordinary matter (that is composed of non-interacting particles with non-zero rest mass, and that behaves as a perfect fluid with zero pressure): for such an ``ordinary dust" we have 
\be\label{T_dust}
T^{\mu \nu } = \rho ^\ast U^\mu U^\nu,
\ee
where $\rho^\ast \geq 0$ is the proper rest-mass density, and where
\be\label{4-velocity}
U^\mu \equiv \dd x^\mu/\dd s
\ee 
is the four-velocity field, so that (\ref{T = tensor product}) applies with $V^\mu \equiv  \sqrt{\rho ^\ast }\, U^\mu$. For ordinary dust we deduce from (\ref{4-velocity}) that the coordinate 3-velocity is
\be
u^i \equiv \frac{\dd x^i}{\dd t} = \frac{\dd x^i}{\dd s}\frac{\dd s}{\dd t} =c\,\frac{U^i}{U^0},
\ee
hence from (\ref{T_dust})
\be\label{veloc_dust}
u^i = c\,\frac{T^{0 i}}{T^{0 0}}.
\ee
The latter relation is not generally true for a continuous medium, e.g. it does not apply to a perfect fluid if the pressure is not zero; simultaneously, in that case, Eq. (\ref{T = tensor product}) is not true either. 
But it is natural from the case of a dust to assume that (\ref{veloc_dust}) applies to a continuous medium when Eq. (\ref{T = tensor product}) is true. If one applies (\ref{veloc_dust}) to the energy-momentum tensor $\Mat{T}_\mathrm{field}$ of an e.m. field for which Eq. (\ref{T = tensor product}) is true, then one finds from $ \mathrm{tr}\, \Mat{T}_\mathrm{field}=0$ that the velocity ${\bf v}_\mathrm{field}$ defined with physical clocks, with components $v_\mathrm{field}^i=\frac{1}{\beta} u^i$, has modulus $c$ as determined with the physical space metric $\Mat{g}$ (\cite{A54}, Eq. (70)):
\be\label{v^2=c^2}
v_\mathrm{field}^2 \equiv \Mat{g}({\bf v}_\mathrm{field},{\bf v}_\mathrm{field}) = c^2.
\ee
In turn, this is consistent with the fact that when (\ref{T = tensor product}) is true we have a ``dust of photons", which behaves so also dynamically \cite{A54} (see Sect. \ref{Solution} below). Moreover, it can be proved that, in order that an e.m. field verify Eq. (\ref{T = tensor product}), it is necessary and sufficient that this be a ``null field", i.e., that both invariants be zero \cite{A54, Stephani1982}: 
\be\label{null field}
{\bf E.B}=0, \qquad{\bf E}^2 = c^2\,{\bf B}^2. 
\ee
In a coordinate system adapted to $\mathcal{E}$ and such that, at the event $X$ considered, we have $\beta (X)=1$ and $g_{ij}(X)=\delta _{ij}$, the tensor $\Mat{F}$ has the form (\ref{F-matrix}). It follows that $T^{0i}$ is given by (\ref{T^0 i}) [in which now the indices of the components $E^i$ and $B^i$ have been lowered with the ``physical" spatial metric $\Mat{g }$], and we get from this by (\ref{veloc_dust}):
\be\label{v_field_null}
{\bf v}_\mathrm{field} = c ({\bf E}\wedge {\bf B})/\abs{{\bf E}\wedge {\bf B}} \equiv c {\bf k}.
\ee
Thus, $v_\mathrm{field}=c$ for an e.m. field whose T-tensor has the form (\ref{T = tensor product}). Surprisingly, {\it for such 
a ``null field",} both $v_\mathrm{field}=c$ and Eq. (\ref{f.(v_field-v_charges)=0}) are true, although $v_\mathrm{charged\ 
medium}<c$ and even $v_\mathrm{charged\ medium}\ll c$ in usual conditions. To see this, remember that ${\bf f} 
\equiv {\bf f}_\mathrm{charged\ medium}$ is the Lorentz force density, given by \be
{\bf f} = \rho _\mathrm{el} \left ({\bf E}+ {\bf v}\wedge {\bf B} \right )
\ee
(now we note again ${\bf v}\equiv {\bf v}_\mathrm{charged\ medium}$ for shortness). We get from this and from (\ref{v_field_null}):
\be\label{f.v}
{\bf f}{\bf .}{\bf v}=\rho _\mathrm{el} {\bf E.v}, \quad {\bf f}{\bf .}{\bf v}_\mathrm{field}=c\rho _\mathrm{el}\left ({\bf v}\wedge {\bf B} \right ){\bf .k}.
\ee
By (\ref{null field}) and (\ref{v_field_null}), we can take (at the event $X$) a spatial basis $({\bf i},{\bf j},{\bf k})$ which is orthonormal for $\Mat{g}$ and such that
\be
{\bf E} = cB {\bf i},\quad {\bf B}= B {\bf j}.
\ee 
Setting ${\bf v}=v_1 {\bf i} + v_2{\bf j} + v_3{\bf k}$, we thus get: 
\be
{\bf E.v} = cBv_1, \quad {\bf v}\wedge {\bf B} = B(v_1 {\bf k}- v_3 {\bf i}), \quad c\left ({\bf v}\wedge {\bf B} \right ){\bf .k}=cBv_1.
\ee
That is, from (\ref{f.v}), we have Eq. (\ref{f.(v_field-v_charges)=0}). Clearly, the foregoing proof depends in an essential way on the assumption that the e.m. field is a null field, so if that is not the case Eq. (\ref{f.(v_field-v_charges)=0}) has no reason to apply. In fact for a general e.m. field it is not clear at all how one should define its velocity field ${\bf v}_\mathrm{field}$.\\

{\bf In summary:} the equation of continuum dynamics (\ref{Eq T-f}) applied to the charged continuum, together with Assumptions \hyperlink{Ass-i}{(i)} and \hyperlink{Ass-ii}{(ii)} above, imply Eq. (\ref{Eq T-f-field}) --- from which the discarded second group (\ref{Eq T-field-3}) follows. Equation (\ref{Eq T-f-field})$_2$ \hypertarget{DynEq_Field}{has the form of the equation of continuum dynamics} (\ref{Eq T-f})$_2$ as applied to the ``field continuum" having the energy-momentum tensor (\ref{T em}), with the force density on the field continuum being the opposite of the Lorentz force exerted on the charged continuum by the field continuum, Eq. (\ref{f_field}). Moreover, Eq. (\ref{Eq T-f-field})$_1$ also has the form of the equation of continuum dynamics (\ref{Eq T-f})$_1$ as applied to the ``field continuum"
, if the velocity of the field continuum is well defined and its projection on the direction of the Lorentz force 
is the same 
as that of the velocity of the charged continuum, Eq. (\ref{f.(v_field-v_charges)=0}). While Eq. (\ref{f.(v_field-v_charges)=0}) turns out to apply in the case of a ``null field", it has no reason to be true for a general e.m. field, for which one does not even know how to define the velocity field ${\bf v}_\mathrm{field}$. So Eq. (\ref{Eq T-f-field})$_1$  has no reason to be true for a general e.m. field.\\

The latter is one conceptual reason, admittedly not very strong, why Eq. (\ref{Eq T-field-3}) is not the right Maxwell second group of the theory --- in addition to the hard fact that it leads to charge production/destruction at high rates. Recall that, at least for a dust, the equation of continuum dynamics (\ref{Eq T-f}) is {\it derived} from Newton's second law, hence it should apply to the charged medium. Thus to avoid Eq. (\ref{Eq T-f-field}) and the discarded second group (\ref{Eq T-field-3}), either \hyperlink{Ass-i}{Assumption (i)} or \hyperlink{Ass-ii}{Assumption (ii)} has to be abandoned. 

\section{The solution of the problem}\label{Solution}

We can't leave \hyperlink{Ass-ii}{Assumption (ii)}, because the concept of a ``total" energy-momentum tensor obeying Eq. (\ref{Eq T}) is necessary to the theory of gravitation. So it is \hyperlink{Ass-i}{Assumption (i)} that has to be abandoned. This means that there must exist an additional energy-momentum-stress tensor, let us call it the ``interaction tensor", such that the total tensor obeying Eq. (\ref{Eq T}) is given by
\be\label{T_with_interact}
\Mat{T} = \Mat{T}_\mathrm{charged\ medium} + \Mat{T}_\mathrm{field} + \Mat{T}_\mathrm{inter}.
\ee
We note that, in general, at an event $X$ for which $\Mat{T}_\mathrm{charged\ medium} \ne {\bf 0}$, we have also $\Mat{T}_\mathrm{field} \ne {\bf 0}$, so that we are in the presence of a {\it mixture} (in the precise sense used in the theory of diffusion): the two constituents of that mixture are the charged medium and the e.m. field. \hypertarget{Mixture}{It is then standard} that indeed the effective energy-momentum-stress tensor of the mixture as a whole is not the sum of the energy-momentum-stress tensors of its constituents \cite{Truesdell1962,Muller1968}. (It thus does not mean that there is an additional medium present beyond the charged medium and the e.m. field.) Now, given the necessary \hyperlink{Ass-ii}{Assumption (ii)} and the equation of continuum dynamics (\ref{Eq T-f}), Eq. (\ref{Eq T-field-3}) is equivalent to the opposite, i.e., to \hyperlink{Ass-i}{Assumption (i)}. This is another conceptual reason why Eq. (\ref{Eq T-field-3}) is not the right Maxwell second group of the theory. \\

With Eq. (\ref{T_with_interact}) replacing Eq. (\ref{T=Tcharges+Tfield}), it is clear that the equation for continuum dynamics (\ref{Eq T-f}) applied to the charged medium, together with \hyperlink{Ass-ii}{Assumption (ii)}, do not imply Eq. (\ref{Eq T-f-field}) any more; hence they do not determine the modified Maxwell second group any more. Therefore, the usual ``gravitationally-modified" second group valid in GR and in the other ``metric theories of gravitation":
\be\label{Maxwell GR}
F^{\mu \nu } _{\ \ \,;\nu  }= -\mu_0 J^\mu,
\ee
is not precluded any more, as it was before in the investigated theory. We show in Appendix \ref{Maxwell GR mean} that Eq. (\ref{Maxwell GR}), as well as the first group (\ref{Maxwell 1}), can be written in terms of the spatial metric and the local time in the synchronized preferred reference frame $\mathcal{E}$, and then take nearly the usual form of the flat-spacetime Maxwell equations for 3-vectors. Thus the standard 2nd group (\ref{Maxwell GR}) is well compatible with the present theory. As is known, it leads to exact charge conservation. At the present stage, other forms of the modified Maxwell second group can not be excluded either, provided they would be found to lead to low-enough charge production in usual situations. We will now show how the compatibility with geometrical optics, which was proved in Sect. 6 of Ref. \cite{A54} with the discarded second group (\ref{Eq T-field-3}), holds true in a more general situation, with emphasis on the case of the standard second group (\ref{Maxwell GR}).\\

The main modification to be made to the argument there, is that one needs to assume that an external force density ${\bf f}_\mathrm{field}$ is indeed acting on the field continuum, as the ``reaction" of the charged medium to the Lorentz force exerted on it by the field continuum. For the link with geometrical optics i.e. with Newton's second law applied to individual photons, we have only to consider the case of a ``null field", i.e. the case that the T-tensor (\ref{T em-dev}) of the e.m. field has the form (\ref{T = tensor product}). In that case, as shown in Ref. \cite{A54}, the {\it spatial} part (\ref{Eq T-f})$_2$ of the continuum dynamical equation is equivalent to Newton's second law for a ``substantial" volume element of the field continuum, in the form (\cite{A54}, Eq. (25) with ${\bf f}'={\bf f}$ for a dust):
\footnote{\
Here the spatial vector {\it gravity acceleration} ${\bf g}$ is given by \cite{A54}
\be \label{g_beta}
\mathbf{g} \equiv -c^2\frac{\mathrm{grad}_{\Mat{g}}\beta}{ \beta}
\ee
with $(\mathrm{grad}_{\Mat{g}}\beta)^i \equiv g^{ij}\beta_{,j}$ where $(g^{ij})$ is the inverse matrix of matrix $(g_{ij})$, and $\frac{D}{Dt}$ is the relevant time derivative ensuring that Leibniz' rule for the derivation of a scalar product is verified \cite{A54,A16}. In terms of the T-tensor, the coordinate velocity ${\bf u}$ is given by (\ref{veloc_dust}).
}
\be\label{Newton 2nd law continuum}
\delta {\bf F} +\frac{\delta E}{c^2} {\bf g} 
=\frac{1}{\beta}\,\frac{D}{Dt}\left (\frac{\delta E}{c^2} {\bf v} \right ), 
\ee
where 
\be\label{delta thing}
\delta {\bf F}\equiv {\bf f} \delta V, \quad \delta E\equiv T^0_{\ \, 0}\delta V, \quad {\bf v}\equiv \frac{1}{\beta }\frac{\dd {\bf x}}{\dd t} \equiv  \frac{1}{\beta } {\bf u}. 
\ee
 This equivalence is valid if the T-tensor has the form (\ref{T = tensor product}), independently of the nature of the continuum. 
\footnote{\
This was noted in Ref. \cite{A54}, Sect. 6, though in the case that the external force density field is ${\bf f}={\bf 0}$. However, the proof given in  Ref. \cite{A54}, Sect. 3.2, applies with the definitions $\rho \equiv T^{\, 00}/(\tilde{\beta }^2 c^{2})$ and $u^{\, i} \equiv c T^{i 0}/T^{00}$ to a medium for which the T-tensor has the form (\ref{T = tensor product}), independently of its nature, provided there is no force internal to this continuum, ${\bf f}={\bf f}'$.
}
Therefore, as noted in Ref. \cite{A54}, it is valid also for a null e.m. field, and this indeed behaves as a ``dust of photons". In Ref. \cite{A54}, it was also proved that the {\it time} part  (\ref{Eq T-f})$_1$ of the continuum dynamical equation is equivalent to the energy equation. The latter is transposed to a continuum from the following form valid for a test particle \cite{A54}:
\be\label{dE/dt} 
\frac{\dd \left( {E{\kern 1pt}\beta } \right)}{\dd t}=E\frac{\partial {\kern 
1pt}\beta }{\partial {\kern 1pt}t} + \beta^{\, 2} {\bf F}{\bf .v}, \qquad
{\bf F}{\bf .v} \equiv  \Mat{g}({\bf F}, 
{\bf v}) \equiv  g_{ij}  F^i v^j.
\ee
While the proof of that equivalence in Ref. \cite{A54} was limited to the case without external force density, i.e. the case that ${\bf f}=0$ in (\ref{Newton 2nd law continuum}), it is straightforward to extend it to show that (\ref{Eq T-f})$_1$ is equivalent to
\be\label{Energy free dust-0}
\frac{\dd \left ( \delta E\, \beta \right )}{\dd t} = \delta E \,\frac{\partial \beta}{\partial t} + \beta ^2 \delta {\bf F.v},
\ee
in which $\delta E$ and $\delta {\bf F}$ are defined by Eq. (\ref{delta thing}).
\footnote{\
By the way, it is not difficult to check that Eqs. (\ref{Newton 2nd law continuum})--(\ref{Energy free dust-0}) are covariant under any change (\ref{change_E}), thus including any change $x'^0=\phi (x^0)$, even though for (\ref{Energy free dust-0}) this has not been noted in Ref. \cite{A54}.

}\\

Thus, when the T-tensor of the e.m. field has the form (\ref{T = tensor product}), the dynamics of a volume element that is followed in its motion is just the same as that of an individual photon subjected to the gravitation and to an external force (detailed in Appendix \ref{DynPhot}), and it may equivalently be defined by the dynamical equation (\ref{Eq T-f}), as applied to the field continuum:
\be\label{Eq T-f-field-1}
T_\mathrm{field \ \,;\nu}^{0 \nu} =b^0(\Mat{T}_\mathrm{field})+\frac{{\bf f}_\mathrm{field}{\bf .v}_\mathrm{field}}{c\beta},
\qquad T_\mathrm{field\ \,;\nu}^{i \nu} =b^i(\Mat{T}_\mathrm{field})+ f_\mathrm{field}^i.
\ee
Note that in the latter equation we have always, in view of (\ref{T em-dev}):
\be\label{Eq T-field-2}
\mu_0 T_{\mathrm{field}\ \,;\nu}^{\mu  \nu}= -F^\mu_{\ \ \lambda }\,F^{\nu \lambda }_{\ \ ;\nu }.
\ee
(This is Eq. (61) of Ref. \cite{A54}, rewritten in the MKSA system.) Therefore, with the standard second group (\ref{Maxwell GR}), we get the well-known equation
\be\label{Eq T-field-4}
 T_{\mathrm{field}\ \,;\nu}^{\mu  \nu}= -F^\mu_{\ \ \lambda }\,J^\lambda.
\ee
{\it In vacuo} ($J^\mu=0$), Eqs. (\ref{Eq T-f-field-1}) and (\ref{Eq T-field-4}) give us:
\be\label{PhotonDust with MaxwellGR}
0=T_\mathrm{field \ \,;\nu}^{0 \nu} =b^0(\Mat{T}_\mathrm{field})+\frac{{\bf f}_\mathrm{field}{\bf .v}_\mathrm{field}}{c\beta},
\qquad 0=T_\mathrm{field\ \,;\nu}^{i \nu} =b^i(\Mat{T}_\mathrm{field})+ f_\mathrm{field}^i.
\ee
This shows that there must indeed be an external force acting on the photon dust (the ``reaction" to the Lorentz force), in addition to the gravitation.  
\section{Discussion}\label{Conclu}

The main conclusions of this work are as follows:\\

(i) The formerly proposed modification of Maxwell's second group in a gravitational field \cite{A54} in the investigated theory 
predicts unrealistically high rates of production/destruction of electric charge. Therefore, that 
gravitationally-modified Maxwell 2nd group has to be discarded.\\

(ii) The theoretical reason for that is the former assumption \cite{A54} according to which the total energy tensor which obeys the dynamical equation in a gravitational field (\ref{Eq T}), is the sum of the energy tensor of the charged medium and that of the e.m. field, Eq. (\ref{T=Tcharges+Tfield}). This assumption is not consistent with the fact that \hyperlink{Mixture}{these two media form a mixture} and, in addition, has a \hyperlink{DynEq_Field}{consequence} which has no reason to be verified in general. \\

(iii) Therefore, one must assume an additional, ``interaction" energy-momentum tensor, such that Eq. (\ref{T=Tcharges+Tfield}) is replaced by Eq. (\ref{T_with_interact}). With this, Maxwell's second group in a gravitational field is less constrained, in particular the standard version (\ref{Maxwell GR}) valid in GR becomes well compatible with the investigated theory.\\

(iv) Also, one must assume that, at least for a null e.m. field (which indeed can be considered as a continuous medium with a well-defined velocity field, and to which one may definitely apply Newton's second law of the present theory), there is a reaction force exerted on the e.m. field by the charged medium.\\

\noi Equation (\ref{T_with_interact}) means that the presence of usual matter producing an e.m. field necessarily gives rise 
(according to the present theory) to the presence of an additional kind of energy, with energy  tensor $\Mat{T}_\mathrm{inter}$. 
The latter does not generally vanish outside the charged medium that emits the e.m. field. If the standard 2nd group 
(\ref{Maxwell GR}) is assumed, we get immediately from (\ref{Eq T-charged}), (\ref{Eq T}), and (\ref{Eq T-field-4}):
\be\label{Dyn-T_Interact with MaxwellGR}
T_\mathrm{inter \ \,;\nu}^{\mu \nu} =b^\mu(\Mat{T}_\mathrm{field})+b^\mu(\Mat{T}_\mathrm{inter}).
\ee
Without the $b^\mu(\Mat{T}_\mathrm{field})$ term, this equation would be identical with the dynamical equation in a gravitational field (\ref{Eq T}), with the energy tensor $\Mat{T}_\mathrm{inter}$ in the place of the total tensor $\Mat{T}$. The time component of Eq. (\ref{Dyn-T_Interact with MaxwellGR}) rewrites as:
 \footnote{\ 
One sees that by using Eqs. (23) and (25) in Ref. \cite{A35} plus Eqs. (\ref{b^mu}) and (\ref{Space metric-v2}) above. The second step uses also Eq. (\ref{W_field}).
 }
\be
\left (T_\mathrm{inter}^{0 0} \right)_{,0} + \left ( T_\mathrm{inter}^{0 j} \right )_{,j} =(\mathrm{Log}\,\beta)_{,0} T_\mathrm{inter}^{0 0} + b^0(\Mat{T}_\mathrm{field}),
\ee
or
\be\label{Balance_W_inter}
\left (T_\mathrm{inter}^{0 0} \right)_{,0} + \left ( T_\mathrm{inter}^{0 j} \right )_{,j} = (\mathrm{Log}\,\beta)_{,0} \left ( T_\mathrm{inter}^{0 0}  -\beta ^{-4} W_\mathrm{field} \right).
\ee
Each of the two source terms on the r.h.s. is proportional to the variation of the gravitational field in the 
preferred 
reference frame: $(\mathrm{Log}\,\beta)_{,0}\simeq -c^{-3}\partial _T U$ in a weak field. However, the first term (with 
$T_\mathrm{inter}^{0 0}$) expresses the usual energy conservation in the investigated theory: the conservation of the total 
energy has just the same form as (\ref{Balance_W_inter}), without the term involving $W_\mathrm{field}$ and with the total 
tensor $\Mat{T}$ in the place of $\Mat{T}_\mathrm{inter}$. That usual energy conservation is a balance between matter energy and 
gravitational energy \cite{A35}. Thus, one may consider that the source of the interaction energy really is the second term that 
is proportional to the e.m. energy $W_\mathrm{field}$, Eq. (\ref{W_field}). The e.m. energy should have  grossly an ellipsoidal 
distribution around the center of a galaxy, which should be much less flat than the mass distribution of the luminous objects, 
since the e.m. energy decreases as $1/r^2$ from each of these objects. Hence we may a priori expect that the same applies to the 
interaction energy $T_\mathrm{inter}^{0 0}$. As is well known, to explain the motions at a galactic scale 
one is led 
to assume a distribution of unseen, ``dark" matter, which should fill a ``halo" around the center of a galaxy or a cluster of galaxies. Therefore, it seems natural to conjecture that the interaction energy, which is necessarily present according to the present theory, is a contribution to dark matter. This is worth a further investigation.

\appendix
\section{Meaning of the standard gravitationally-modified Maxwell equations}\label{Maxwell GR mean}

First, let us observe that, with the definitions (\ref{E and B}) for the electric and magnetic fields, Maxwell's {\it first} group (\ref{Maxwell 1}) can be rewritten almost exactly in the usual form of the Maxwell-Gauss and Maxwell-Faraday equations, namely:
\be\label{Homogen Maxwell}
\mathrm{div}_\Mat{g}\, {\bf B} =0,\qquad \mathrm{rot}_\Mat{g}\, {\bf E} = -\frac{\partial {\bf B}}{\partial t_{\bf x}},
\ee 
the mere difference being thus in the use of the local time (\ref{dt_x}) (i.e., $\frac{\partial }{\partial t_{\bf x}} =\frac{1}{\beta (t,{\bf x})} \frac{\partial }{\partial t}$) of the synchronized reference frame $\mathcal{E}$ (i.e., $\gamma _{0 i}=0$) and in the fact that the operators are defined with the help of the spatial metric $\Mat{g}$ in the frame $\mathcal{E}$:
\be\label{div_g & rot_g}
\mathrm{div}_\Mat{g} {\bf B} \equiv B^i_{\mid i} = \frac{1}{\sqrt{g}} \left( \sqrt{g} B^i \right )_{,i},\qquad \left(\mathrm{rot}_\Mat{g} {\bf E} \right )^i \equiv e^{i j}_{\ \ k} E^k_{\mid j},
\ee
with $_{\mid j}$ the covariant derivative associated with $\Mat{g}$. Indeed, as one easily checks from the definition (\ref{E and B}),  Eq. (\ref{Homogen Maxwell}) coincides with (\ref{Maxwell 1}) in coordinates $x^\mu $ such that, at the event $X$ considered, we have
\be\label{Cart loc g & beta=1}
g_{ij}(X) = \delta _{ij}, \quad g_{ij,k}(X)=0, \quad \beta(X) =1
\ee
 (see Eq. (24.14) of Fock \cite{Fock1964}). Starting from one coordinate system that is adapted to the preferred frame and that verifies the synchronization condition $\gamma _{0 i}=0$, one can get to another one that in addition verifies (\ref{Cart loc g & beta=1}), by a change
\be\label{change_E}
x'^i =\psi ^i(x^1,x^2,x^3),\qquad x'^0=\phi (x^0).
\ee
Since each of the two equations in (\ref{Homogen Maxwell}) is invariant under such a change, our statement is proved. In nearly the same way, from the relation valid in any coordinates for the antisymmetric tensor $F^{\mu \nu }$:
\be
F^{\mu \nu } _{\ \ \,;\nu  } = \frac{1}{\sqrt{-\gamma }} \left ( \sqrt{-\gamma } F^{\mu \nu} \right )_{,\nu},
\ee
and using Eq. (\ref{gamma}) for $\gamma $, we find that the $\mu =0$ component of Eq. (\ref{Maxwell GR}) [the standardly-modified second group] rewrites as the Maxwell-Poisson equation in terms of metric $\Mat{g}$:
\be\label{Poisson}
\mathrm{div}_\Mat{g} {\bf E} = \mu _0 c^2 \rho_\mathrm{el} \equiv \frac{\rho_\mathrm{el}}{\epsilon _0},
\ee
whereas the spatial components of Eq. (\ref{Maxwell GR}) rewrite as a space vector equation involving an additional term as compared with the flat-spacetime Maxwell-Amp\`ere equation:
\be\label{Maxwell-Ampere}
\mathrm{rot}_\Mat{g}\, {\bf B} -\frac{1}{c^2} \frac{\partial {\bf E}}{\partial t_{\bf x}} - \frac{1}{c^2} \,{\bf g}\wedge {\bf B} =\mu _0 {\bf j},
\ee
where $j^i\equiv J^i$ and the spatial vector ${\bf g}$ is given by Eq. (\ref{g_beta}). To rewrite (\ref{Maxwell GR}) as (\ref{Poisson}) and (\ref{Maxwell-Ampere}), we use coordinates that, in addition to (\ref{Cart loc g & beta=1}), are such that 
\be\label{d beta/dt =0}
(\partial \beta /\partial x^0)(X)=0.
\ee
The full set (\ref{Cart loc g & beta=1}) and (\ref{d beta/dt =0}) can be fulfilled by a change (\ref{change_E}); cf. Ref. \cite{L&L}, around Eq. (85.18). Equations (\ref{Poisson}) and (\ref{Maxwell-Ampere}) are invariant under a change (\ref{change_E}). Note that the derivation applies in any synchronized reference frame, but the gravity acceleration vector (\ref{g_beta}) makes little sense in a general situation unless one assumes the preferred-frame dynamics of the investigated theory.

\section{Dynamics of a photon under gravitational and non-gravitational forces}\label{DynPhot}

Our extension of Newton's second law has exactly the same form for a mass particle and for a photon, i.e. \cite{A54,A16,A15}
\be\label{Newton 2nd law}
{\bf F} + \frac{E}{c^2} {\bf g} = \frac{1}{c^2}\frac{D(E{\bf v})}{Dt_{\bf x}} \equiv \frac{1}{c^2\,\beta }\frac{D(E{\bf v})}{Dt}.
\ee
For a photon, we define $E = h\nu $, $h$ being Planck's constant and $\nu $ the frequency as measured with the local time: $\nu\equiv \dd n/\dd t_{\bf x} \equiv (1/\beta )\dd n/\dd t$ with $n$ the number of periods. The energy equation derived from (\ref{Newton 2nd law}) is also the same for a mass particle and for a photon, i.e., Eq. (\ref{dE/dt}). In the case without an external force ${\bf F}$, the proof has been given in full for a mass particle in Ref. \cite{A15}, and has been outlined also in Ref. \cite{A15} for a photon. Here we give the proof with ${\bf F}$ for a photon, for completeness. Equation (\ref{Newton 2nd law}) is equivalent to:
\be\label{Newton 2nd law-2}
E {\bf g} + c^2 {\bf F}= E \frac{D{\bf v}}{Dt_{\bf x}} + \frac{\dd E}{\dd t_{\bf x}} \,{\bf v},
\ee
whence by taking the scalar product $\Mat{g}$ with ${\bf v}$, using Leibniz' rule verified \cite{A16,A15} by the $D/Dt$ derivative:
\be
E {\bf g.v} = E \frac{\dd }{\dd t_{\bf x}} \left( \frac{{\bf v}^2}{2} \right ) + \frac{\dd E}{\dd t_{\bf x}} {\bf v}^2 - c^2 {\bf F.v}.
\ee
Since ${\bf v}^2 =c^2$ for a photon and since ${\bf v}\equiv (1/\beta ){\bf u}$, this rewrites as
\be\label{E g.u}
\frac{E}{\beta} {\bf g.u} = c^2 \left ( \frac{\dd E}{\dd t_{\bf x}}-{\bf F.v}\right ).
\ee
From (\ref{g_beta}), we have
\be\label{g.u}
 {\bf g.u} = -c^2\frac{\Mat{g}(\mathrm{grad}_{\Mat{g}}\beta,{\bf u})}{\beta} = -\frac{c^2}{\beta} \beta _{,i} \frac{\dd x^i}{\dd t}= -\frac{c^2}{\beta} \left (\frac{\dd \beta }{\dd t} - \frac{\partial  \beta }{\partial  t} \right ).
\ee
Hence, (\ref{E g.u}) is
\be\label{E g.u-2}
-E\left (\frac{\dd \beta }{\dd t} - \frac{\partial  \beta }{\partial  t} \right ) = \beta  \frac{\dd E}{\dd t}-\beta ^2{\bf F.v},
\ee
whence follows Eq. (\ref{dE/dt}).\\

Thus, for a photon, to deduce the energy equation (\ref{dE/dt}) from Newton's second law (\ref{Newton 2nd law}), we used the property ${\bf v}^2 =c^2$. Now we show that the energy equation allows us to rewrite Newton's second law in a form which ensures that  ${\bf v}^2 =c^2$ is indeed maintained at all times. Equation (\ref{dE/dt}) is equivalent to
\be
\beta \frac{\dd E}{\dd t} +E\left (\frac{\partial  \beta }{\partial  t}+\beta _{,i} u^i \right )= E \frac{\partial  \beta }{\partial  t} +\beta ^2 \,{\bf F.v}.
\ee
Removing the term present on both sides and using again (\ref{g.u}), this gives us
\be\label{dE/dt-2}
\frac{\dd E}{\dd t} = \beta  \left ({\bf F.v}+ \frac{E}{c^2} {\bf g.v} \right ).
\ee
With this, Newton's second law (\ref{Newton 2nd law-2}) rewrites as
\be\label{Newton 2nd law-3}
E \frac{D{\bf v}}{Dt_{\bf x}} + \left ({\bf F.v}+ \frac{E}{c^2} {\bf g.v} \right ) {\bf v} = E {\bf g} + c^2 {\bf F},
\ee
or
\be
\frac{D{\bf v}}{Dt_{\bf x}} = {\bf g}- \left( \frac{{\bf g.v}}{c^2} + \frac{{\bf F.v}}{E} \right ) {\bf v} + \frac{c^2}{E} {\bf F}.
\ee
It follows from the latter equation that
\be
\frac{\dd }{\dd t_{\bf x}} \left( \frac{{\bf v}^2}{2} \right )={\bf v.}\frac{D{\bf v}}{Dt_{\bf x}} =(c^2-{\bf v}^2) \left ( \frac{{\bf g.v}}{c^2} +\frac{{\bf F.v}}{E} \right ).
\ee
Note that all of this is true for a mass particle as well as for a photon. Thus, if we have ${\bf v}^2 =c^2$ at the initial time, this condition is maintained at all times. In general relativity, the condition ${\bf v}^2 =c^2$ (i.e., $\dd s^2=0$) for a photon is not dynamically implied by the (geodesic) law of motion, instead it is assumed from the outset as one considers a ``null geodesic".

\end{document}